\title{Classification of integrable Volterra type lattices on the sphere. Isotropic case}
\author{V.E. Adler\thanks{L.D. Landau Institute for Theoretical Physics,
 1a Semenov pr., 142432 Chernogolovka, Russia.\newline
 E-mail: {\tt adler@itp.ac.ru}}}
\date{17 December 2007}
\documentclass[a4paper,12pt]{article}
\usepackage{amsmath,amssymb,amsthm}
\usepackage[english]{babel}
\usepackage{hyperref}

\advance\textwidth 1in
\advance\oddsidemargin -0.5in
\advance\textheight 1.2in
\advance\topmargin -0.6in

\theoremstyle{plain}
\newtheorem{theorem}{Theorem}
\newtheorem{statement}{Statement}
\theoremstyle{remark}
\newtheorem{remark}{Remark}

\def\a{\alpha}
\def\b{\beta}
\def\g{\gamma}
\def\d{\delta}
\def\eps{\varepsilon}
\def\la{\lambda}
\def\SP<#1>{\langle#1\rangle}
\def\Complex{\mathbb C}
\def\const{\mathop{\rm const}}
\def\Im{\mathop{\rm Im}}
\def\wedgec{\underset{\text{'}}{\wedge}}
\def\wedgec{\wedge\mspace{-12.5mu}\text{\raisebox{-0.2em}[0em][0em]{,}}\mspace{8mu}}

\begin{document}\maketitle

\begin{abstract}
The symmetry approach is used for classification of integrable isotropic
vector Volterra lattices on the sphere. The list of integrable lattices
consists mainly of new equations. Their symplectic structure and associated
PDE of vector NLS-type are discussed.
\end{abstract}

\section{Introduction}

We call vector Volterra lattices the equations of the general form
\begin{equation}\label{Vnx}
 V_{n,x}=f_nV_{n+1}+g_nV_n+h_nV_{n-1},\quad n\in{\mathbb Z}
\end{equation}
where $V_n$ are vectors and $f_n,g_n,h_n$ are scalar functions depending on
$V_{n+1},V_n,V_{n-1}$. The integrability is understood as existence of
higher symmetries, that is the equations which are consistent with
(\ref{Vnx}), but involve the larger number of neighbor vectors (preserving
the same quasi-linear structure). The precise definitions are given in next
Section. The goal of this paper is to classify integrable cases under the
following assumptions:

(i) the lattice and its symmetries are isotropic and shift invariant, that
is their coefficients depend only on the scalar products
$v_{m,n}:=\SP<V_m,V_n>=\SP<V_n,V_m>$ and this dependence is the same at each
node;

(ii) the lattice must be integrable independently on the dimension of the
vector space and the nature of scalar product;

(iii) all $V_n$ are of unit length, $v_{n,n}=1$.

The shift invariance allows the use of shorthand notation with the discrete
variable $n$ omitted from subscripts, so that equation (\ref{Vnx}) takes the
form
\begin{equation}\label{Vx}
 V_x=fV_1+gV+hV_{-1}
\end{equation}
(subscripts $x,t$ will be reserved for denoting derivatives, not shifts).
Due to the other assumptions, functions $f,g,h$ are related by equation
\begin{equation}\label{fgh}
 v_{1,0}f+g+v_{0,-1}h=0
\end{equation}
and depend only on the scalar products $v_{1,0},v_{0,-1},v_{1,-1}$ which can
be considered as independent variables. Therefore, the classification
problem is reduced to finding of two functions of three variables, so that
its complexity is comparable with the case of scalar Volterra lattices
\begin{equation}\label{vx}
 v_x=f(v_1,v,v_{-1})
\end{equation}
classified by Yamilov \cite{Yamilov_1983}, see also the recent review
article \cite{Yamilov_2006}. The whole method of solution is also very
close, since the necessary integrability conditions in both cases formally
coincide (the difference is in the set of dynamical variables: $v_{m,n}$
instead of $v_n$). In the continuous case, the general approach based on
this remarkable observation has been developed by Sokolov and Meshkov in the
pioneering papers \cite{Meshkov_Sokolov_2002,Meshkov_Sokolov_2004} devoted
to the classification of KdV-type vector equations (including the
anisotropic ones) on the sphere. Important classification results for some
other types of vector PDE were obtained in \cite{Sokolov_Wolf_2001,
Tsuchida_Wolf, Anco_Wolf}, however the approach in these papers relied
essentially on the polynomial or rational structure of equations.

In principle, the classification problem for the lattices (\ref{Vx}) can be
solved without the unitary condition (iii). This constraint does not define
an independent class of equations, but only a special reduction of the
general problem. Indeed, it can be resolved by use of the stereographic
projection
\[
 V=\frac{1-\SP<U,U>}{1+\SP<U,U>}\,e_0+\frac{2}{1+\SP<U,U>}U,
\]
where $e_0$ is some fixed unit vector and $U$ belongs to its orthogonal
subspace. Vector $U$ satisfies, in virtue of equation (\ref{Vx}), some
isotropic lattice $U_x=\tilde fU_1+\tilde gU+\tilde hU_{-1}$. Since the
dimension of the vector space is inessential in our considerations, we see
that any lattice on the sphere corresponds under this mapping to some
lattice in the free space. On the other hand, this lattice for $U$ is not
arbitrary: it must admit the reduction $\SP<U,U>=1$ since $U=V$ under this
constraint. This reduction brings back to the original lattice.

The paper is organized as follows. Section \ref{s:cond} contains a concise
explanation of symmetry approach and derivation of the sequence of
integrability conditions in the form of conservation laws. These are used in
Section \ref{s:classification} which is the main technical body of the text.
All lattices (\ref{Vx}) are divided there into two subclasses; the first one
is analyzed thoroughly, while the second one is poor in answers and its
presentation is more brief. The results of classification are presented in
Section \ref{s:list}. The rest of the paper contains some discussion of
associated PDEs and symplectic structures.

\section{The necessary integrability conditions}\label{s:cond}

The symmetry approach to classification of integrable equations had been
developed in 80's, see e.g. \cite{Sokolov_Shabat_1984,
Mikhailov_Shabat_Yamilov_1987} as general sources, \cite{Levi_Yamilov_1997}
for a modern account on the discrete case and review articles
\cite{VAdler_Shabat_Yamilov,Yamilov_2006} for detailed references. The
lattice (\ref{Vx}) is called integrable if it possesses an infinite
hierarchy of the symmetries of the form
\begin{equation}\label{Vtk}
 V_{t_k}=p^{(k,k)}V_k+p^{(k,k-1)}V_{k-1}+\dots
 +p^{(k,1-k)}V_{1-k}+p^{(k,-k)}V_{-k}
\end{equation}
with coefficients depending on the scalar products of $V_k,\dots,V_{-k}$. It
is easy to see that the compatibility condition splits over the vector
variables $V_n$ resulting in the commutator relation
\begin{equation}\label{FP}
 D_x(P^{(k)})-D_{t_k}(F)=[F,P]
\end{equation}
for {\em scalar} operators
\[
 F=fT+g+hT^{-1},\quad P^{(k)}=p^{(k,k)}T^k+\dots+p^{(k,-k)}T^{-k}
\]
where $T$ denotes the shift operator $n\mapsto n+1$. This allows to use the
necessary integrability conditions established in the scalar case (\ref{vx})
by Yamilov \cite{Yamilov_1983}, with operator $F$ instead of the
linearization operator $f_*=f_{v_1}T+f_v+f_{v_{-1}}T^{-1}$. For sake of
completeness we repeat very briefly the derivation of these conditions.

Equation (\ref{FP}) is equivalent to a set of equations for the coefficients
of $P^{(k)}$. One pair of equations defines explicitly the leading
coefficients
\[
 p^{(k,k)}=f_{k-1}\dots f_1f,\quad
 p^{(k,-k)}=\a h_{-k+1}\dots h_{-1}h,
\]
while solvability of the rest equations provides some sequence of necessary
conditions to integrability of the lattice. These conditions do not depend
actually on the order $k$ of the symmetry. More precisely, let equation
(\ref{FP}) can be solved, at some $k=K$, with respect to $2l$ coefficients
$p^{(k,\pm k)},p^{(k,\pm(k-1))},\dots,p^{(k,\pm(k-l))}$, where $k-l>1$. Then
it can be solved with respect to these $2l$ coefficients at any $k>K$.
Moreover, the coefficients of one symmetry are expressed through the
coefficients of the other one by explicit formulae. In order to prove this,
it is sufficient to notice that the term $D_{t_k}(F)$ in the l.h.s. of
(\ref{FP}) affects the computation of the coefficients $p^{(k,1)},
p^{(k,0)},p^{(k,-1)}$ only, and that the special form of the leading
coefficients written above allows to approximate $P^{(k)}$ by the formal
power series $(P^{(K)})^{k/K}$. This brings to the following statement.

\begin{statement}
If the lattice (\ref{Vx}) possesses an infinite hierarchy of higher
symmetries then the equations
\begin{align*}
 & L_x=[F,L],\quad L=a^{(-1)}T+a^{(0)}+a^{(1)}T^{-1}+a^{(2)}T^{-2}\dots\\
 & \tilde L_x=[F,\tilde L],\quad
   \tilde L=\tilde a^{(-1)}T^{-1}+\tilde a^{(0)}
   +\tilde a^{(1)}T+\tilde a^{(2)}T^2\dots
\end{align*}
are solvable with respect to the coefficients $a^{(j)}$, $\tilde a^{(j)}$
depending on $v_{m,n}$.
\end{statement}

The series $L,\tilde L$ are called {\em formal symmetries}. In turn, the
equations for their coefficients can be rewritten further as the sequence of
conservation laws
\begin{equation}\label{rs}
 D_x(\rho^{(j)})=(T-1)(\sigma^{(j)}),\quad
 D_x(\tilde\rho^{(j)})=(T^{-1}-1)(\tilde\sigma^{(j)}),\quad j=0,1,2,\dots
\end{equation}
More precisely, if the lattice (\ref{Vx}) possesses the symmetry of order
$k$, then equations (\ref{rs}) can be solved with respect to
$\sigma^{(j)},\tilde\sigma^{(j)}$ for $j=0,\dots,k-2$. The densities
$\rho^{(j)},\tilde\rho^{(j)}$ are expressed explicitly by certain recursive
algorithm in terms of the lattice coefficients and previously found
$\sigma^{(j)},\tilde\sigma^{(j)}$. This algorithm relates $\rho^{(j)}$ with
the residue of $L^j$ defined as the free term of power series in $T$ (the
formula $\mathop{\rm res}[A,B]\in\Im(T-1)$ can be proven). However, in
practice we will need only few several conservation laws and the
corresponding formulae can be derived straightforwardly.

\begin{statement}\label{st:rs}
Let the lattice (\ref{Vx}) be integrable, then equations (\ref{rs}) are
solvable for the following sequence of the densities $\rho^{(j)}$,
$\tilde\rho^{(j)}$:
\begin{align}
 \label{rho0}
 \rho^{(0)}&=\log f,& \tilde\rho^{(0)}&=\log h, \\
 \label{rho1}
 \rho^{(1)}&=g+\sigma^{(0)},& \tilde\rho^{(1)}&=g+\tilde\sigma^{(0)},\\
 \label{rho2}
 \rho^{(2)}&=hf_{-1}+\frac12(\rho^{(1)})^2+\sigma^{(1)},&
 \tilde\rho^{(2)}&=fh_1+\frac12(\tilde\rho^{(1)})^2+\tilde\sigma^{(1)}.
\end{align}
\end{statement}
\begin{proof}
The equations for the coefficients $a^{(-1)},a^{(0)},a^{(1)},a^{(2)}$ are:
\begin{align*}
   0&=fa^{(-1)}_1-f_1a^{(-1)},\\
 a^{(-1)}_x&=fa^{(0)}_1-fa^{(0)}+ga^{(-1)}-g_1a^{(-1)},\\
 a^{(0)}_x&=fa^{(1)}_1-f_{-1}a^{(1)}+ha^{(-1)}_{-1}-h_1a^{(-1)},\\
 a^{(1)}_x&=fa^{(2)}_1-f_{-2}a^{(2)}
  +ga^{(1)}-g_{-1}a^{(1)}+ha^{(0)}_{-1}-ha^{(0)}.
\end{align*}
The first equation implies $a^{(-1)}=f$, without loss of generality. Then
the second equation takes the form $(\log f)_x=(T-1)(a^{(0)}-g)$, so that we
obtain the density $\rho^{(0)}$ and the formula for the next coefficient of
the formal symmetry: $a^{(0)}=g+\sigma^{(0)}$. Accordingly to the third
equation, this coefficient may be taken as the density $\rho^{(1)}$ and then
$a^{(1)}=h+\sigma^{(1)}/f_{-1}$. The last equation can be brought to the
form
\[
 \Bigl(hf_{-1}+\frac12(\rho^{(1)})^2+\sigma^{(1)}\Bigr)_x
  =(T-1)(f_{-1}f_{-2}a^{(2)}+\sigma^{(1)}\rho^{(1)}_{-1})
\]
after multiplication by $f_{-1}$ and taking into account the previous
equations. The second set of the densities is obtained immediately due to
the symmetry $n\to-n$.
\end{proof}

\begin{remark}
In addition to the higher symmetries, existence of the higher order
conservation laws is another characteristic feature of integrable equations.
It is possible to derive some integrability conditions from this property as
well. This leads to the notion of {\em formal conservation law}
\[
 S_x+SF+F^\top S=0,\quad S=s^{(0)}+s^{(1)}T^{-1}+s^{(2)}T^{-2}+\dots
\]
where $(aT^j)^\top:=T^{-j}a$ and coefficients $s^{(j)}$ depend on $v_{m,n}$.
Solvability of this equation is equivalent to the sequence of conditions of
the form
\begin{equation}\label{^rs}
 \hat\rho^{(j)}=(T-1)(\hat\sigma^{(j)}),\quad j=0,1,2,\dots
\end{equation}
In particular,
\[
 \hat\rho^{(0)}=\log(-f/h),\quad \hat\rho^{(1)}=2g+D_x(\hat\sigma^{(0)}).
\]
It can be proven that conservation laws (\ref{rs}) are equivalent in virtue
of conditions (\ref{^rs}), that is
$\rho^{(j)}+\const\tilde\rho^{(j)}\in\Complex\oplus\Im(T-1)$. In some
classification problems use of these additional integrability conditions may
lead to a crucial simplification or even to a shorter list of equations. In
particular, these conditions were used by Yamilov in his classification of
the scalar lattices (\ref{vx}) (see footnote on p.~567 and Theorem 22 in
\cite{Yamilov_2006}). It turns out, however, that in the vector case these
conditions are of minimal value and it is possible to dispense with them (in
all found lattices they are fulfilled automatically).
\end{remark}

Returning to the characteristic equation (\ref{FP}) we notice that
solvability of the first pair of integrability conditions (\ref{rs}),
(\ref{rho0}) allows to find the coefficients $p^{(k,\pm k)}$,
$p^{(k,\pm(k-1))}$ of the symmetry. At $k=2$ this defines the symmetry
completely, due to the constraint $\SP<V,V>=1$ which implies
\[
 v_{2,0}p^{(2,2)}+v_{1,0}p^{(2,1)}+p^{(2,0)}
 +v_{0,-1}p^{(2,-1)}+v_{0,-2}p^{(2,-2)}=0.
\]
The straightforward computation shows that if this symmetry exists then it
must be of the form
\begin{equation}\label{Vt}
\begin{aligned}
 V_t&=ff_1(V_2-v_{2,0}V)+f(\rho^{(1)}_1+\rho^{(1)})(V_1-v_{1,0}V)\\
 &\qquad +\kappa h(\tilde\rho^{(1)}_{-1}+\tilde\rho^{(1)}+\tilde\kappa)
  (V_{-1}-v_{0,-1}V)+\kappa hh_{-1}(V_{-2}-v_{0,-2}V)
\end{aligned}
\end{equation}
with some indeterminate integration constants $\kappa,\tilde\kappa$.
Although the use of this explicit formula gives no essential advantage in
solving the classification problem, it is useful as a final check of
integrability of the obtained lattices.

\section{Analysis of the integrability conditions}\label{s:classification}

\subsection{First step}\label{s:step1}

Consider the first pair of integrability conditions (\ref{rs}), (\ref{rho0})
\begin{equation}\label{cond.1}
 D_x(\log f)\in\Im(T-1),\quad D_x(\log h)\in\Im(T-1).
\end{equation}
It is easy to obtain the following equations as a corollary:
\begin{equation}\label{cond.1:3}
 \frac{f_{v_{1,-1}}}{f^2}+\frac{h}{f}T\genfrac{(}{)}{}{}{f_{v_{1,-1}}}{f^2}=0,\quad
 \frac{h_{v_{1,-1}}}{h^2}+\frac{f}{h}T^{-1}\genfrac{(}{)}{}{}{h_{v_{1,-1}}}{h^2}=0.
\end{equation}
Indeed, the terms containing scalar products $v_{k,k-3}$ appear only by
differentiating $v_{1,-1}$ with respect to $x$:
\begin{gather*}
 D_x(\log f)=\frac{f_{v_{1,-1}}}{f}D_x(v_{1,-1})+\dots=
 \frac{f_{v_{1,-1}}}{f}(f_1v_{2,-1}+h_{-1}v_{1,-2})+\dots\\
 \stackrel{\Im(T-1)}{\simeq}
 \left(\frac{f_{v_{1,-1}}}{f}f_1
  +T\genfrac{(}{)}{}{}{f_{v_{1,-1}}}{f}h\right)v_{2,-1}+\dots
\end{gather*}
and the first equation (\ref{cond.1:3}) follows. This computation is
actually equivalent to applying of {\em variational derivative}
$\d/\d_{v_{3,0}}$ defined by formula
\[
 \frac{\d a}{\d v_{j,0}}= \frac{\partial}{\partial v_{j,0}}
  \sum^\infty_{k=-\infty}T^k(a),\quad j=1,2,\dots
\]
The use of this notion makes the computations more algorithmic, due to the
equality
\[
 \Complex\oplus\Im(T-1)=\bigcap^\infty_{j=1}\ker\frac{\d}{\d v_{j,0}}
\]
which is proven along the same lines as in scalar case \cite{Yamilov_2006}.

\begin{statement}\label{th:case12}
The dependence of the coefficients of the lattice on $v_{1,-1}$ may be one
of the following:
\begin{align*}
 \text{Case 1.}&\qquad f=\frac{a(v_{0,-1})}{v_{1,-1}+b(v_{1,0},v_{0,-1})},\quad
  h=-\frac{a(v_{1,0})}{v_{1,-1}+b(v_{1,0},v_{0,-1})},\\
 \text{Case 2.}&\qquad f=f(v_{1,0},v_{0,-1}),\quad h=h(v_{1,0},v_{0,-1}).
\end{align*}
\end{statement}
\begin{proof}
First equation (\ref{cond.1:3}) implies that $f_{v_{1,-1}}/f^2$ may depend
on $v_{0,-1}$ only. If $f_{v_{1,-1}}\ne0$ then we come to the Case 1. If
$f_{v_{1,-1}}=0$ then $h_{v_{1,-1}}=0$ as well, in virtue of the second
equation (\ref{cond.1:3}), and we come to the Case 2.
\end{proof}

Conditions (\ref{cond.1}) are far from being exhausted by this statement. We
will see that in the Case 1 they allow to define functions $a$, $b$ as well.

\subsection{Case 1: $f_{v_{1,-1}}\ne0$}\label{s:case1}

Notice that in this case the relation (\ref{^rs}) at $j=0$ is satisfied with
$\hat\sigma^{(0)}=-\log a(v_{0,-1})$. This means that conditions
(\ref{cond.1}) are equivalent to each other and we may consider only the
first one. Applying of $\d/\d_{v_{2,0}}$ to it is a rather tedious task. The
resulting equation is polynomial in variables $v_{k+2,k}$ and vanishing of
the coefficients brings to a certain overdetermined system for functions $a$
and $b$. It is convenient to introduce the auxiliary functions
\begin{equation}\label{ab0}
 y(v)=\frac{1-v^2}{a^2(v)},\quad c(u,v)=\frac{b(u,v)+uv}{a(u)a(v)}
\end{equation}
and to denote $u=v_{1,0}$, $v=v_{0,-1}$, $w=v_{-1,-2}$. This allows to
rewrite the system in a relatively compact form as follows:
\begin{align}
\label{ab1}
 & c(u,v)(a'(u)-a'(v))=(a(u)y(u))_u-(a(v)y(v))_v,\\
\label{ab2}
 & a(u)(c+y(u))c_u-a(v)(c+y(v))c_v=
   \frac{u(c-y(v))}{a(u)}-\frac{v(c-y(u))}{a(v)},\quad c=c(u,v),\\
\label{ab3}
 & (c(v,w)+y(v))(2c(u,v)+y(v))_v=(c(u,v)+y(v))(2c(v,w)+y(v))_v.
\end{align}
At first, we will prove that all solutions of equation (\ref{ab3}) are:
\begin{align*}
 &\text{(i)}\qquad 2c(u,v)=2\a-y(u)-y(v),\\
 &\text{(ii)}\qquad c(u,v)=\a z(u)z(v)+\b,\quad y(v)=\g z^2(v)-\b,\quad
 z'\ne0
\end{align*}
where $\a,\b,\g$ are arbitrary constants.

If $c(v,w)+y(v)=0$ or $c(u,v)+y(v)=0$ then (\ref{ab3}) is reduced to the
equation
\[
 0=(y(u)-y(v))y'(v),
\]
hence $y(v)=-\b$, $c(u,v)=\b$, a special case of solution (ii).

If $(c(v,w)+y(v))(c(u,v)+y(v))\ne0$ then the variables in (\ref{ab3}) can be
separated:
\begin{equation}\label{ab3'}
  \frac{(2c(u,v)+y(u))_u}{c(u,v)+y(u)}=2k(u),\quad
  \frac{(2c(u,v)+y(v))_v}{c(u,v)+y(v)}=2k(v)
\end{equation}
and as a corollary we obtain $c_{uv}=k(u)c_v=k(v)c_u$. The case $k=0$
corresponds to the solution (i). At $k\ne0$ we get $c=C(K(u)+K(v))$, $K'=k$,
$C''=C'$, whence $c=\a z(u)z(v)+\b$, where $z'=kz$. Moreover, both equations
(\ref{ab3'}) are reduced to the relation
\[
 y'(v)=\frac{2z'(v)}{z(v)}(y(v)+\b)
\]
and we get (ii) by integration. Now we consider both types of solutions
separately and come to the following statement.

\begin{statement}
The solutions $a=a(v)$, $b=b(u,v)$ of the system (\ref{ab0})--(\ref{ab3})
are exhausted, up to the scaling $a\to\const a$, by the following list:
\begin{align}
\label{ab.s1}
 & a=v-1/v,\quad b=-uv;\\[0.5em]
\label{ab.s2}
 & a^2-kva+v^2-1=0,\quad b=a(u)a(v)-uv;\\[0.5em]
\label{ab.s3}
 & a=v+\eps,\quad b=-1;\\
\label{ab.s4}
 & a=v+\eps,\quad b=(u+\eps)(v+\eps)
  \left(\sqrt{\Bigl(\frac{u-\eps}{u+\eps}-k\Bigr)
              \Bigl(\frac{v-\eps}{v+\eps}-k\Bigr)}+k\right)-uv;\\
\label{ab.s5}
 & a=v+\eps,\quad b=1+\eps(u+v)+k\sqrt{(u+\eps)(v+\eps)}
\end{align}
where $\eps=\pm1$ and $k$ is an arbitrary constant.
\end{statement}

\begin{proof}
{\em Solutions of type} (i). Applying $\partial_u\partial_v$ to (\ref{ab1})
yields
\[
 y'(u)a''(v)=y'(v)a''(u).
\]
If $y'=0$ then scaling allows to set $y=1$, $a^2(v)=1-v^2$ and then
(\ref{ab1}) implies that $c=1$. Equation (\ref{ab2}) becomes identically
true in virtue of these relations and we arrive to solution (\ref{ab.s2}) at
$k=0$.

If $y'\ne0$ then $a'=\mu y+\nu$. The variables in (\ref{ab1}) are now
separated and we obtain the overdetermined ODE system for the functions
$a=a(v)$, $y=y(v)$:
\begin{equation}\label{ab:ay}
 ay'=R(y)=-\frac32\mu y^2+(\a\mu-\nu)y+\la,\quad a'=S(y)=\mu y+\nu,\quad ya^2=1-v^2.
\end{equation}
Differentiation yields
\[
 a(2yS+R)=-2v,\quad S(2yS+R)+(2S+2y\dot S+\dot R)R+2=0.
\]
The polynomial on $y$ in the l.h.s.\! of the latter equation must vanish
identically since $y'\ne0$. This gives the relations $\mu=0$, $\la\nu=-1$
and moreover, the scaling allows to set $\nu=1$. Now, the system
(\ref{ab:ay}) is reduced to equations
\[
 ay'=-y-1,\quad a=v+\eps,\quad a^2y=1-v^2.
\]
It is easy to prove that they are consistent at $\eps^2=1$, and an
intermediate substitution into (\ref{ab2}) proves that $\a=0$. The resulting
solution is (\ref{ab.s3}).
\smallskip

{\em Solutions of type} (ii). Applying $\partial_u\partial_v$ to (\ref{ab1})
yields
\begin{equation}\label{az}
 \a\left(a'(u)-a'(v)+\frac{z(u)a''(u)}{z'(u)}-\frac{z(v)a''(v)}{z'(v)}\right)=0.
\end{equation}
If $\a=0$ then $c=\b$ and variables in equation (\ref{ab2}) are separated:
\[
\frac{(\b-y(u))a(u)}{u}=\frac{(\b-y(v))a(v)}{v}=\d.
\]
This relation turn the equation (\ref{ab1}) into identity as well. Taking
(\ref{ab0}) into account, we obtain the equation $\b a^2-\d va+v^2-1=0$ for
$a(v)$. This brings, up to the scaling, to the solutions (\ref{ab.s1}),
(\ref{ab.s2}).

If $\a\ne0$ then we set $\a=1$ without loss of generality. Equation
(\ref{az}) implies $a'=\mu/q+\nu$, then the variables in (\ref{ab1}) are
separated and we obtain the overdetermined ODE system for the functions
$a=a(v)$, $z=z(v)$:
\begin{equation}\label{az1}
 a'=\frac{\mu}{z}+\nu,\quad
 ((\g z^2-\b)a)'-\frac{\mu\b}{z}+\mu z=\la,\quad
 (\g z^2-\b)a^2=1-v^2.
\end{equation}
Notice that $\g\ne0$: otherwise $-2\mu\b/z+\mu z-\b\nu=\la$ and since
$z'\ne0$, hence $\mu=0$; but then the equations $a'=\nu$, $\b a^2=v^2-1$ are
inconsistent. Therefore, second equation (\ref{az1}) can be rewritten as
follows:
\[
 z'=\frac{1}{2\g a}\left(
  -\g\nu z-\mu(\g+1)+\frac{\la+\b\nu}{z}+\frac{2\mu\b}{z^2}\right).
\]
Now, differentiating of third equation (\ref{az1}) brings, as in the
previous case, to a polynomial equation for $z$ which must be satisfied
identically. This gives equations for the parameters:
\[
 (\g-1)\b\mu=0,\quad (3\g-1)(\la+\b\nu)\mu=0,\quad (\g-3)\mu\nu=0,\quad
 4\g(\la\nu+1)+(\g-1)^2\mu^2=0.
\]
Moreover, substitution into (\ref{ab2}) gives additionally the equations
\[
 (\g+1)(\g-3)\b\mu=0,\quad (\g^2-1)(\la+\b\nu)=0,\quad (\g^2-1)\mu=0.
\]
The solutions of the whole system are:
\begin{gather*}
 (\mu^2=1,\b=0,\la=0,\nu=0,\g=-1),\\
 (\mu=0,\nu=-1/\la,\g^2=1),\quad
 (\mu=0,\nu=-1/\la,\b=\la^2).
\end{gather*}
The first one is unsuitable since it leads to $z'=0$. For the other two we
set $\nu=1$, $\la=-1$, $a=v+\eps$ without loss of generality. It is easy to
check that (\ref{az1}) are consistent at $\eps^2=1$ and we come to solutions
(\ref{ab.s4}) and (\ref{ab.s5}), respectively.
\end{proof}

It can be proved straightforwardly that conditions (\ref{rs}) at $j=0$ are
fulfilled for each solution (\ref{ab.s1})--(\ref{ab.s5}), that is there
exist quantities $\sigma^{(0)}$, $\tilde\sigma^{(0)}$ which turn them into
identities. It is sufficient to compute only $\sigma^{(0)}$, due to the
relation $\tilde\sigma^{(0)}_{-1}=D_x(\hat\sigma^{(0)})-\sigma^{(0)}$ where
$\hat\sigma^{(0)}=-\log a(v_{0,-1})$. Practically, this computation is based
on the ``summation by parts'' algorithm, see e.g. \cite[Theorem
1]{Yamilov_2006}. After finding $\sigma^{(0)}$ one can continue the
integrability test with the next pair of densities (\ref{rho1}). It turns
out that in all cases except for (\ref{ab.s5}) the second integrability
condition is fulfilled automatically. In the case (\ref{ab.s5}) we obtain
the restriction $k^3-4k=0$ on the values of parameter. In more details, the
density $\rho^{(1)}$ is in this case of the form
\begin{align*}
 \rho^{(1)}=~&\frac{f_{-1}}{v_{-1,-2}+\eps}(v_{0,-2}-1)
   +\frac{ff_{-1}}{v_{-1,-2}+\eps}\Bigl(
             v_{1,-2}-v_{1,0}+v_{0,-1}-v_{-1,-2} \\
  & -\frac12\left(k\sqrt{v_{1,0}+\eps}+2\eps\sqrt{v_{0,-1}+\eps}\right)
            \left(k\sqrt{v_{-1,-2}+\eps}+2\eps\sqrt{v_{0,-1}+\eps}\right)\Bigr)
\end{align*}
and it can be proven that $\d D_x(\rho^1)/\d_{v_{2,0}}$ vanishes if and only
if the above constraint holds.

The computation of $\sigma^{(1)}$ and further check of the integrability
conditions require the considerable efforts. Fortunately, it is possible to
avoid these calculations by checking that the explicit formula (\ref{Vt})
provide the higher symmetry indeed. This turns out to be true for
(\ref{ab.s1})--(\ref{ab.s4}) and (\ref{ab.s5}) at $k=0,\pm2$ (with constants
$\kappa=-1$, $\tilde\kappa=0$ in all cases) and we come, respectively, to
the lattices (\ref{V1})--(\ref{V5}) in the List \ref{list} below.

\subsection{Case 2: $f_{v_{1,-1}}=0$}\label{s:case2}

Computations here are easier, but also more lengthy, since in some subcases
we have to check up to three integrability conditions (\ref{rs}). However,
the result of this search is somewhat disappointing: it consists of one
lattice (\ref{V6}). By this reason we give only schematic account of this
case.

Applying $\d/\d_{v_{2,0}}$ to (\ref{cond.1}) yields the equations
\begin{equation}\label{hf}
\begin{aligned}
 & \frac{h}{f}\left(T\genfrac{(}{)}{}{}{f_{v_{0,-1}}}{f}
       +\frac{f_{v_{1,0}}}{f}\right)+
   \frac{f_{v_{0,-1}}}{f}+T^{-1}\genfrac{(}{)}{}{}{f_{v_{1,0}}}{f}=0,\\
 & \frac{h}{f}\left(T\genfrac{(}{)}{}{}{h_{v_{0,-1}}}{h}
       +\frac{h_{v_{1,0}}}{h}\right)+
   \frac{h_{v_{0,-1}}}{h}+T^{-1}\genfrac{(}{)}{}{}{h_{v_{1,0}}}{h}=0.
\end{aligned}
\end{equation}
In turn, differentiating this with respect to $v_{2,1}$ yields
\[
 (\log f)_{v_{1,0},v_{0,-1}}=0,\quad (\log h)_{v_{1,0},v_{0,-1}}=0
 \quad\Rightarrow\quad f=T(a)b,\quad h=T(c)d
\]
where $a,b,c,d$ are functions on $v_{0,-1}$. Now, the variables in equations
(\ref{hf}) are separated and we come to relations
\[
 \frac{(ab)'}{ab}\cdot\frac{c}{a}=\mu,\quad
 \frac{(ab)'}{ab}\cdot\frac{b}{d}=-\mu,\quad
 \frac{(cd)'}{cd}\cdot\frac{b}{d}=\nu,\quad
 \frac{(cd)'}{cd}\cdot\frac{c}{a}=-\nu
\]
with some constants $\mu,\nu$. If $ab+cd\ne0$ then $(ab)'=(cd)'=0$, so that
two cases are possible, up to the scaling:
\begin{align*}
 &\text{(i)}\qquad b=p/a,\quad
   c=ap/p',\quad d=-p'/a,\quad p'\ne0;\\
 &\text{(ii)}\qquad a=\a/b,\quad d=1/c.
\end{align*}
In the case (i), applying $\d/\d_{v_{1,0}}$ to (\ref{cond.1}) brings to
certain overdetermined system for functions $a,p$. It is convenient to
analyze this system taking into account some additional information (namely,
the equation $pp''=\const(p')^2$) which can be obtained either from the
integrability condition (\ref{^rs}) at $j=1$ or from the next pair of
conservation laws (\ref{rs}), (\ref{rho1}). This allows to prove that
functions $a(v),p(v)$ may be the following:
\[
 a=p=\frac{1}{v+\d};\qquad a=1,~p=v+\d;\qquad a=v,~p=v^3.
\]
The check of conservation laws (\ref{rs}), (\ref{rho1}) for the first
solution proves that $\d$ must take the values $\pm1,0$ and leads to the
lattice (\ref{V6}), while two other solutions do not pass the test.

In the case (ii) the first pair of integrability conditions (\ref{rs}),
(\ref{rho0}) is fulfilled for any $\a,b,c$. The further analysis proves that
conditions (\ref{rs}), (\ref{rho1}) are fulfilled if $\a=1$ and either
$b(v)=c(v)=\sqrt{v+\d}$ or $b=c=1$. However, the next conditions (\ref{rs}),
(\ref{rho2}) fail in both cases, so that this case turns out to be empty.

\section{The list of integrable lattices}\label{s:list}

\def\mylistname{List} \let\oldtablename=\tablename \let\tablename=\mylistname

\begin{table}[t]
\hrulefill
\begin{align}
\label{V1}\tag{$V_1$}
 & V_x=\frac{a(V_1-v_{1,0}V)+a_1(v_{0,-1}V-V_{-1})}
   {v_{1,-1}-v_{1,0}v_{0,-1}}\,,\quad a=v_{0,-1}-\frac{1}{v_{0,-1}};\\[0.5em]
\label{V2}\tag{$V_2$}
 & V_x=\frac{a(V_1-v_{1,0}V)+a_1(v_{0,-1}V-V_{-1})}
   {v_{1,-1}-v_{1,0}v_{0,-1}+aa_1}\,,\quad
a^2-2kv_{0,-1}a+v^2_{0,-1}-1=0;\\[0.5em]
\label{V3}\tag{$V_3$}
 & V_x=\frac{(v_{0,-1}+\eps)(V_1+\eps V)-(v_{1,0}+\eps)(V_{-1}+\eps V)}
            {v_{1,-1}-1}\,;\\[0.5em]
\label{V4}\tag{$V_4$}
 & V_x=\frac{(v_{0,-1}+\eps)(V_1+\eps V)-(v_{1,0}+\eps)(V_{-1}+\eps V)}
  {v_{1,-1}-v_{1,0}v_{0,-1}+(v_{1,0}+\eps)(v_{0,-1}+\eps)(k+pp_1)}\,,
 \quad p=\sqrt{\frac{v_{0,-1}-\eps}{v_{0,-1}+\eps}-k};\\[0.5em]
\label{V5}\tag{$V_5$}
 & V_x=\frac{(v_{0,-1}+\eps)(V_1+\eps V)-(v_{1,0}+\eps)(V_{-1}+\eps V)}
  {v_{1,-1}+\eps(v_{1,0}+v_{0,-1})+1
   +k\sqrt{v_{1,0}+\eps}\sqrt{v_{0,-1}+\eps}}\,,
  \quad k=0,\pm2;\\[0.5em]
\label{V6}\tag{$V_6$}
 & V_x=\frac{V_1+\d V}{v_{1,0}+\d}-\frac{V_{-1}+\d V}{v_{0,-1}+\d}
 \,,\quad\d=0,\pm1.
\end{align}
\caption{Integrable lattices, $\SP<V,V>=1$,
 $v_{m,n}=\SP<V_m,V_n>$, $\eps=\pm1$.}
\label{list}
\hrulefill
\end{table}
\let\tablename=\oldtablename

\begin{theorem}
If isotropic Volterra type lattice on the sphere $\SP<V,V>=1$ satisfies
integrability conditions (\ref{rs})--(\ref{rho2}) then it coincides with one
of the lattices from the List \ref{list}, up to scaling of $x$. Each lattice
from this list possesses at least one higher symmetry of the form
(\ref{Vt}).
\end{theorem}

\begin{remark}
The lattices corresponding to the different signs of $\eps$ or $\d$ are
equivalent modulo flip map $V_n\to(-1)^nV_n$. The lattice (\ref{V2}) at
$k=\pm1$ coincides with (\ref{V5}) at $k=0$.
\end{remark}

The lattice (\ref{V6}) is the discrete Heisenberg spin chain introduced in
\cite{Ragnisco_Santini}, see also \cite{Bobenko,Bobenko_Suris} where the
applications to the discrete geometry were considered and \cite{Adler_2000}
where the anisotropic version (see Section \ref{s:remarks}) was studied. It
can be written (at $\d=1$ and after scaling $x$) as
\begin{equation}\label{Heisenberg}
 V_x=\frac{V_1+V}{|V_1+V|^2}-\frac{V+V_{-1}}{|V+V_{-1}|^2}\,.
\end{equation}
In this form, the constraint $\SP<V,V>=1$ is not necessary for
integrability. This lattice and its higher symmetry (\ref{Vt}) can be
written compactly as
\[
 V_x=(T-1)(W),\quad V_t=(T-1)P_W(W_1-W_{-1}),\quad W=(V+V_{-1})^{-1}
\]
by use of the operations
\[
 A^{-1}=\frac1{\SP<A,A>}A,\qquad P_A(B)=2\SP<A,B>A-\SP<A,A>B.
\]
The variable $U$ satisfies the polynomial lattices
\[
 W_x=-P_W(W_1-W_{-1}),\quad
 W_t=-P_W(P_{W_1}(W_2+W)-P_{W_{-1}}(W+W_{-2}))
\]
which are integrable not only in the vector case, but also in more general
setting related to Jordan triple systems \cite{Adler_Svinolupov_Yamilov}.

The lattices (\ref{V1})--(\ref{V5}) are new, up to the author's knowledge.
The lattice (\ref{V3}) is related to (\ref{V6}) by composition of difference
substitution and reduction. Namely, first we can resolve the constraint
$\SP<V,V>=1$ by use of stereographic projection as explained in
Introduction. This brings (\ref{V3}) at $\eps=-1$ to the form
\[
 U_x=\frac{|U-U_{-1}|^2(U_1-U)+|U_1-U|^2(U-U_{-1})}{|U_1-U_{-1}|^2}
\]
and then substitution $\tilde V=U-U_{-1}$ brings it to the lattice
\[
 \tilde V_x=
   \frac{|\tilde V|^2\tilde V_1+|\tilde V_1|^2\tilde V}
        {|\tilde V_1+\tilde V|^2}
  -\frac{|\tilde V_{-1}|^2\tilde V+|\tilde V|^2\tilde V_{-1}}
       {|\tilde V+\tilde V_{-1}|^2}\,.
\]
This is not the same lattice as (\ref{Heisenberg}), however it is obvious
that both lattices admit the reduction on sphere which brings them to the
lattice (\ref{V6}). The question on the substitutions for the other lattices
from the list is so far open.

\section{Associated partial differential equations}\label{s:PDE}

The very general observation due to Levi \cite{Levi_1981} is that a higher
symmetry of an integrable lattice gives rise to some PDE after elimination of
the discrete variable $n$. The lattice itself is now interpreted as B\"acklund
transformation for this PDE. The examples of such relation can be found in
\cite{Shabat_Yamilov_1988,Shabat_Yamilov_1991} and many other works. In
particular, the integrable Volterra lattices (\ref{vx}) are associated with some
systems of nonlinear Schr\"odinger type. There are known also many results on
the multifield analogs of NLS-type systems, see e.g. \cite{Svinolupov_1992, 
Sokolov_Svinolupov_1994, Svinolupov_Sokolov_1996, Habibullin_Sokolov_Yamilov},
however their classification is far from being completed. The list of vector
Volterra lattices provides several new examples of such systems.

The elimination of the discrete variable is done as follows. The equations 
(\ref{Vx}), (\ref{fgh}) imply the corollaries
\begin{align*}
 \SP<V_x,V_1>&=(1-v^2_{1,0})f+(v_{1,-1}-v_{1,0}v_{0,-1})h,\\
 \SP<V_x,V_x>&=(1-v^2_{1,0})f^2+2(v_{1,-1}-v_{1,0}v_{0,-1})fh+(1-v^2_{0,-1})h^2.
\end{align*}
We assume that these equations can be solved with respect to the scalar
products $v_{1,-1}$, $v_{0,-1}$ (this is true for all lattices from the List
\ref{list}). Then equation (\ref{Vx}) can be rewritten in the form
\begin{equation}\label{V-1}
 V_{-1}=\tilde fV_1+\tilde gV+\tilde hV_x
\end{equation}
with coefficients depending on the scalar products of vectors $V_1,V,V_x$.
Analogously,
\[
 V_2=\hat fV_{1,x}+\hat gV_1+\hat hV.
\]
Iteration of these formulae allows to express all vectors $V_n$ through the
vectors $U=V_1$, $V$ and their derivatives. As a result, the symmetry
(\ref{Vt}) gives rise to a system of the form
\begin{equation}\label{UVt}
\left\{\begin{aligned}
  U_t&=U_{xx}+\a U_x+\b V_x+\g U+\d V,\\
 -V_t&=V_{xx}+\tilde\a U_x+\tilde\b V_x+\tilde\g U+\tilde\d V,
\end{aligned}\right.\qquad \SP<U,U>=\SP<V,V>=1
\end{equation}
with coefficients depending on the scalar products of $U$, $U_x$, $V$ and
$V_x$. The equation (\ref{V-1}) becomes an explicit B\"acklund
auto-transformation
\[
 U_{-1}=V,\quad V_{-1}=\tilde fU+\tilde gV+\tilde hV_x
\]
of this system. Converse is not true: not any integrable system (\ref{UVt})
admits auto-BT of such form. Classification problem for this type of
equations may be difficult, since even the simplest lattices from our list
correspond to rather cumbersome systems (\ref{UVt}). Few instances are given
below. In the case (\ref{V6}) at $\d=\pm1$ we come to the system
\begin{align*}
 U_t &= U_{xx}-\frac{2\SP<U_x,V>+4\d}{\SP<U,V>+\d}U_x
  +\frac{2V_x}{\SP<U,V>+\d}+\left(\frac{\SP<U_x,U_x>}{\SP<U,V>+\d}
        -\frac{2\SP<U,V_x>}{(\SP<U,V>+\d)^2}\right)(\d U+V),\\
-V_t &= V_{xx}-\frac{2\SP<U,V_x>-4\d}{\SP<U,V>+\d}V_x
  -\frac{2U_x}{\SP<U,V>+\d}+\left(\frac{\SP<V_x,V_x>}{\SP<U,V>+\d}
        +\frac{2\SP<U_x,V>}{(\SP<U,V>+\d)^2}\right)(U+\d V),
\end{align*}
while (\ref{V6}) at $\d=0$ corresponds to the system
\begin{align*}
 U_t &= U_{xx}-\frac{2\SP<U_x,V>\SP<U,V>+2}{\SP<U,V>^2}U_x
  +\left(\SP<U_x,U_x>+\frac{2\SP<U_x,V>}{\SP<U,V>}\right)U
  +\genfrac{(}{)}{}{}{2V}{\SP<U,V>}_x,\\
-V_t &= V_{xx}-\frac{2\SP<U,V_x>\SP<U,V>-2}{\SP<U,V>^2}V_x
  +\left(\SP<V_x,V_x>-\frac{2\SP<U,V_x>}{\SP<U,V>}\right)V
  -\genfrac{(}{)}{}{}{2U}{\SP<U,V>}_x.
\end{align*}
The lattice (\ref{V3}) is associated with the system
\begin{align*}
 U_t &= U_{xx} -2\left(\frac{\SP<U,V_x>\SP<U_x,V>}{(\SP<U,V>+\eps)^2}
       -\frac{\SP<U_x,V_x-V>}{\SP<U,V>+\eps}\right)U_x
       -\frac{\SP<U_x,U_x>}{\SP<U,V>+\eps}V_x\\
 &\qquad +\frac{\SP<U_x,U_x>}{\SP<U,V>+\eps}
   \left(1+\frac{\SP<U,V_x>}{\SP<U,V>+\eps}\right)(\eps U+V),\\
-V_t &= V_{xx} +2\left(\frac{\SP<U,V_x>\SP<U_x,V>}{(\SP<U,V>+\eps)^2}
       -\frac{\SP<V_x,U_x+U>}{\SP<U,V>+\eps}\right)V_x
       +\frac{\SP<V_x,V_x>}{\SP<U,V>+\eps}U_x \\
 &\qquad +\frac{\SP<V_x,V_x>}{\SP<U,V>+\eps}
   \left(1-\frac{\SP<U_x,V>}{\SP<U,V>+\eps}\right)(U+\eps V).\\
\end{align*}

\section{Presymplectic structure}\label{s:Hamiltonian}

The bi-Hamiltonian structure of the scalar Volterra lattice is well known,
see e.g. \cite{TF}. In the vector case the question is more difficult and it
requires further investigation. However, the following statement shows that
all lattices under scrutiny possess at least some uniform presymplectic
structure.

\begin{statement}
Any lattice (\ref{V1})--(\ref{V6}) can be written in presymplectic form
\begin{equation}\label{SVx}
 SV_x=\frac{\d H}{\d V}+\la V,\quad H=\rho^{(0)}=\log f(v_{1,-1},v_{1,0},v_{0,-1})
\end{equation}
where $S$ is a certain skew-symmetric operator of the form
\begin{equation}\label{S}
 S=pT^{-1}-p_1T-qV_{-1}V^\top T^{-1}+q_1V_1V^\top T
   +r(V_1V^\top_{-1}-V_{-1}V^\top_1),
\end{equation}
$\la$ is Lagrange multiplier corresponding to the constraint $\SP<V,V>=1$
and operator $UV^\top$ acts accordingly to the formula
$UV^\top(W)=U\SP<V,W>$.
\end{statement}
\begin{proof}
The equation (\ref{SVx}) is equivalent to
\begin{gather*}
 (pT^{-1}-p_1T)(fV_1+gV+hV_{-1})-V_{-1}(qT^{-1}+r)(f+v_{1,0}g+v_{1,-1}h)\\
  +V_1(r+q_1T)(v_{1,-1}f+v_{0,-1}g+h)-\la V\\
 = T\left(\frac{f_{v_{1,-1}}}{f}V_1+\frac{f_{v_{0,-1}}}{f}V\right)
  +\frac{f_{v_{1,0}}}{f}V_1+\frac{f_{v_{0,-1}}}{f}V_{-1}
  +T^{-1}\left(\frac{f_{v_{1,0}}}{f}V+\frac{f_{v_{1,-1}}}{f}V_{-1}\right).
\end{gather*}
Equating the coefficients at $V,V_{\pm2}$ yields
\[
 \la=pf_{-1}-p_1h_1,\quad p=-f_{v_{1,-1}}/f^2,\quad pf+p_1h=0.
\]
The first two equations are just definitions of $\la$ and $p$ while the
latter one is fulfilled for the lattices from the list in virtue of
(\ref{cond.1:3}). Equations for the rest coefficients give the system for
$q$ and $r$ of the form
\begin{equation}\label{qr}
 Ar+A_1q_1=C,\quad Br+B_{-1}q=D
\end{equation}
where
\begin{gather*}
 A=v_{1,-1}f+v_{0,-1}g+h,\quad B=f+v_{1,0}g+v_{1,-1}h,\\
 C=p_1g_1+(\log f_1f)_{v_{1,0}},\quad D=pg_{-1}-(\log ff_{-1})_{v_{0,-1}}.
\end{gather*}
Elimination of one of the unknown functions, say $r$, brings (\ref{qr}) to
the form
\[
 (T-1)(AB_{-1}q)=BC-AD.
\]
This means that the system (\ref{qr}) is solvable if and only if
$BC-AD\in\Im(T-1)$. Remarkably, this condition is equivalent exactly to
$D_x(\log f)\in\Im(T-1)$, as an easy check proves, and therefore it is true
for all lattices from the List \ref{list}.
\end{proof}

The concrete expressions for the coefficients $q,r$ may be rather cumbersome
(it is clear from the proof that they are related somehow with the quantity
$\sigma^{(0)}$). The answer is very simple for the lattice (\ref{V3}):
\begin{equation}\label{pqr3}
 p=\frac{1}{v_{0,-1}+\eps},\quad q=\frac{1}{(v_{0,-1}+\eps)^2},\quad r=0.
\end{equation}
The formula $\SP<U,SW>=\Omega(U,W)$ relates operator $S$ with 2-form
\begin{gather*}
 \Omega=\sum_n\bigl(p_n\SP<dV_n\wedgec dV_{n-1}>
 +q_n\SP<V_n,dV_{n-1}>\wedge\SP<V_{n-1},dV_n>\\
 +r_n\SP<V_{n+1},dV_n>\wedge\SP<V_{n-1},dV_n>\bigr)
\end{gather*}
where $\SP<\a\wedgec\b>(U,W):=\SP<\a(U),\b(W)>-\SP<\a(W),\b(U)>$. It is easy
to see that this form is exact in the case (\ref{pqr3}), namely
$\Omega=d\sum_np_n\SP<V_n,dV_{n-1}>$. Therefore $d\Omega=0$, that is
operator $S$ is symplectic indeed. Unfortunately, this is not true in the
general case.

It is also worth to notice that the representation (\ref{SVx}) can be
replaced with a linear pencil by assuming that Hamiltonian is of the form
$H=\rho^{(0)}+\kappa\rho$, where $\rho$ is some additional conserved density
depending on $v_{1,0}$ (it does not belong to the sequence (\ref{rs}),
however it turns out that such densities exist for all lattices under
consideration). Operator $S$ also acquires linear dependence on $\kappa$,
preserving the same structure (\ref{S}). We bring the explicit formulae only
for the relatively simple case of lattice (\ref{V1}):
\begin{gather*}
 \rho^{(0)}=\log\frac{a}{v_{1,-1}-v_{1,0}v_{0,-1}},\quad
 \rho=\log v_{1,0},\quad
 p=\frac{1}{a},\quad a=v_{0,-1}-\frac{1}{v_{0,-1}},\\[0.3em]
 q=\frac{1}{a^2}
  +(\kappa-1)\frac{(v_{1,-1}-v_{1,0}v_{0,-1})(v_{0,-2}-v_{0,-1}v_{-1,-2})}
    {av_{0,-1}\left(v_{1,-1}-\dfrac{v_{0,-1}}{v_{1,0}}\right)
     \left(v_{0,-2}-\dfrac{v_{0,-1}}{v_{-1,-2}}\right)},\\
 r=\frac{1}{a_1a}+(\kappa-1)\frac{v_{1,-1}-v_{1,0}v_{0,-1}}
   {\left(v_{1,-1}-\dfrac{v_{0,-1}}{v_{1,0}}\right)
    \left(v_{1,-1}-\dfrac{v_{1,0}}{v_{0,-1}}\right)}.
\end{gather*}
Operator $S$ is not symplectic here. We see also that its simplest form
corresponds to the Hamiltonian $\rho^{(0)}+\rho$ rather than $\rho^{(0)}$,
but this may be not so for the other lattices.

\section{Concluding remarks}\label{s:remarks}

The goal of the present paper was to solve some classification problem; such
important things as difference substitutions, Lax pairs, B\"acklund
transformations, explicit solutions and so on have not been considered.
These open problems require, probably, more individual investigation for
each member of the obtained list. From the author's point of view, the
question on the Hamiltonian properties of the vectorial equations is among
the most intriguing ones.

It was mentioned in Introduction that the assumption (iii) can be removed by
use of stereographic projection. Another interesting setting is related with
the variables on the cone $\SP<V,V>=0$ instead of the sphere. At first
sight, this constraint may be treated as a limiting case, but actually it
defines some independent class of equations. In particular, in this case the
coefficient $g$ is not expressed through $f,h$ and we also have no explicit
formula like (\ref{Vt}) for the symmetry. An interesting example here is the
lattice
\[
 V_x=\frac{1}{v_{1,-1}}(v_{0,-1}V_1-v_{1,0}V_{-1})
  +b(v_{1,-1},v_{1,0},v_{0,-1})V, \quad v_{n,n}=0.
\]
It is likely that it satisfies the infinite sequence of integrability
conditions (\ref{rs}) at arbitrary $b$, but (local) symmetries exist only if
$b_{v_{1,-1}}=0$.

The other possible generalizations are related with the condition (i). The
simplest anisotropic lattice is analog of (\ref{V6})
\[
 V_x=\SP<V,KV>\left(\frac{V_1+V}{1+\SP<V_1,V>}
 -\frac{V+V_{-1}}{1+\SP<V,V_{-1}>}\right),\quad \SP<V,V>=1
\]
where $K$ is an arbitrary symmetric operator. This lattice is closely
related to many other integrable equations, among them Sklyanin lattice and
Landau-Lifshitz equation \cite{Adler_2000}. The classification problem in
the anisotropic case can be in principle solved along the same lines (cf
\cite{Meshkov_Sokolov_2002,Meshkov_Sokolov_2004} in the continuous case),
however technically it is much more difficult since coefficients acquire
dependence on the additional variables $\tilde v_{m,n}=\SP<V_m,KV_n>$. It is
interesting to consider also the asymmetric scalar product ($v_{m,n}\ne
v_{n,m}$), however the examples of this type are not known at the moment.

\paragraph{Acknowledgements.} This work was supported by the Russian Foundation
for Basic Researches under grant \# 06-01-92051-KE-a.


\end{document}